\documentclass[12pt]{article}
\usepackage{amssymb}
\usepackage{amsmath}
\usepackage{color}

\usepackage{graphicx}
 
\ifx\pdfoutput\undefined
     \relax
\else
     \usepackage{epstopdf}
\fi

\makeatletter
\def\numberbysection{\@addtoreset{equation}{section}
 	\def\theequation{\thesection.\arabic{equation}}}
\makeatother

\numberbysection

\newcommand{\be}{\begin{eqnarray}}
\newcommand{\ee}{\end{eqnarray}}
\newcommand{\non}{\nonumber}
\newcommand{\tr}{\mathop{\rm tr}\nolimits}

\newcommand{\ch}{\mathop{\rm ch}\nolimits}
\newcommand{\sh}{\mathop{\rm sh}\nolimits}

\def\cP{{\cal P}}   
\def\cS{{\cal S}}   

\newcommand{\half}{\frac{1}{2}}

\definecolor{brique}{rgb}{.9,.2,0}
\definecolor{blvert}{rgb}{0,.8,.85}
\definecolor{vertcl}{rgb}{0,1,.7}
\newcommand\vertcl[1]{\textcolor{vertcl}{#1}}
\newcommand\blvert[1]{\textcolor{blvert}{#1}}
\newcommand\brique[1]{\textcolor{brique}{#1}}
\def\lapth{
\begin{picture}(136,70)(0,-15)\thicklines
\put(0,0){\vertcl{\rule{20pt}{4pt}}}
\put(19,1){\vertcl{\line(1,3){23}}} 
\put(20,1){\vertcl{\line(1,3){23}}} 
\put(21,1){\vertcl{\line(1,3){23}}}
\put(22,1){\vertcl{\line(1,3){23}}}
\put(45,70){\vertcl{\line(1,-3){23}}} 
\put(44,70){\vertcl{\line(1,-3){23}}} 
\put(43,70){\vertcl{\line(1,-3){23}}}
\put(42,70){\vertcl{\line(1,-3){23}}}
\put(2,24){\vertcl{\rule{120pt}{4pt}}}
\put(65,0){\vertcl{\rule{60pt}{4pt}}}
\put(5,37){\Huge{\brique{\textbf{L}}}} 
\put(62,37){\Huge{\brique{\textbf{PTh}}}}
\put(12,-8){\blvert{\rule{92pt}{3.5pt}}}
\put(24,-15){\blvert{\rule{57pt}{3.5pt}}}
\put(36,-22){\blvert{\rule{30pt}{3.5pt}}}
\end{picture}
\raisebox{35pt}{
\begin{minipage}{320pt}\begin{center}
\textbf{Laboratoire d'Annecy-leVieux de Physique
Th\'eorique}\\[4ex]
website: \texttt{http://lappweb.in2p3.fr/lapth-2005/}
\end{center}
\end{minipage}}\\
\vspace{10pt}\quad \hrulefill\\
\vspace{10pt}}

\begin{document}

\pagestyle{empty}
\setcounter{page}{0}
\hspace{-1cm}\lapth

\vfill\vfill

\begin{center}
{\LARGE Complete Bethe Ansatz solution\\
       of the open spin-$s$ XXZ chain\\[1.2ex]
       with general integrable boundary terms}\\

\vfill\vfill

\large Luc Frappat \footnote{
       Laboratoire d'Annecy-le-Vieux de Physique Th\'eorique
       LAPTH, UMR 5108, CNRS and Universit\'e de Savoie, B.P. 110, F-74941 Annecy-le-Vieux 
       Cedex, France},
       Rafael I. Nepomechie \footnote{
       Physics Department, P.O. Box 248046, University of Miami,
       Coral Gables, FL 33124 USA}
   and Eric Ragoucy ${}^{1}$ \\

\end{center}

\vfill\vfill

\begin{abstract}
    We consider the open spin-$s$ XXZ quantum spin chain with $N$
    sites and general integrable boundary terms for generic values of
    the bulk anisotropy parameter, and for values of the boundary
    parameters which satisfy a certain constraint.  We derive two sets
    of Bethe Ansatz equations, and find numerical evidence that
    together they give the complete set of $(2s+1)^{N}$ eigenvalues of
    the transfer matrix.  For the case $s=1$, we explicitly determine
    the Hamiltonian, and find an expression for its eigenvalues in
    terms of Bethe roots.
\end{abstract}

\vfill

\begin{center}
\end{center}

\vfill
\rightline{UMTG--254}
\rightline{LAPTH-1190/07}
\rightline{June 2007}

\setcounter{footnote}{0}
\pagestyle{plain}

\section{Introduction}\label{sec:intro}

Much of the impetus for the development of the so-called Quantum
Inverse Scattering Method and Quantum Groups came from the effort to
formulate and solve higher-spin generalizations of the spin-1/2 XXZ
(anisotropic Heisenberg) quantum spin chain with periodic boundary
conditions \cite{ZF}-\cite{spinsXXZ}.  (See also \cite{BP} and
references therein.)  For the open spin-1/2 XXZ chain \cite{ABBBQ},
Sklyanin \cite{Sk} discovered a commuting transfer matrix based on
solutions of the boundary Yang-Baxter equation (BYBE) \cite{Ch}.  This
made it possible to generalize some of the above works to open spin
chains \cite{MNR}-\cite{FLSU}.  Although general (non-diagonal)
solutions of the BYBE were eventually found \cite{dVGR,GZ,IOZ}, Bethe
Ansatz solutions were known only for open spin chains with diagonal
boundary terms.  Further progress was made in \cite{CLSW, Ne}, where a
solution was found for the open spin-1/2 XXZ chain with general
integrable (i.e., including also non-diagonal) boundary terms,
provided the boundary parameters satisfy a certain constraint.  It was
subsequently realized \cite{NR} that two sets of Bethe Ansatz
equations (BAEs) were generally needed in order to obtain the complete
set of eigenvalues.  (Recently, the second set of BAEs was derived
within the generalized algebraic Bethe Ansatz approach of \cite{CLSW}
by constructing \cite{YZ1} a suitable second reference state.  For
other related work, see e.g. \cite{Do1}-\cite{MNS}, and references
therein.)

We present here a Bethe Ansatz solution of the $N$-site open
spin-$s$ XXZ quantum spin chain with general integrable boundary
terms, following an approach which was developed for the spin-1/2 XXZ
chain in \cite{YNZ}, and then generalized to the spin-1/2 XYZ chain in
\cite{YZ3}.  Doikou solved a special case of this model (among others)
in \cite{Do2} using the method in \cite{Ne}, and also computed some
of its thermodynamical properties.  She further investigated this
model in \cite{Do3} using the method in \cite{CLSW}.  We find a
generalization of Doikou's constraint on the boundary parameters, as
well as a second set of BAEs, which we argue is necessary to obtain
the complete set of $(2s+1)^{N}$ eigenvalues.

The outline of this paper is as follows.  In Sec.  \ref{sec:transfer}
we review the construction of the model's transfer matrices, and the
so-called fusion hierarchy which they satisfy.  In Sec.  \ref{sec:BA},
we identify (as in \cite{YNZ}) the $Q$ operator with a transfer matrix
whose auxiliary-space spin tends to infinity, and obtain a $T-Q$
equation for the model's fundamental (spin-1/2 auxiliary space)
transfer matrix.  Using some additional properties of this transfer
matrix, we arrive at a pair of expressions for its eigenvalues in
terms of roots of corresponding BAEs, as well as a corresponding
constraint on the boundary parameters.  We provide numerical evidence
of completeness of this solution for small values of $s$ and $N$.  In
Sec.  \ref{sec:Ham} we explicitly determine the Hamiltonian for the
case $s=1$, and find an expression for its eigenvalues in terms of 
Bethe roots.  We end in Sec.  \ref{sec:conclude} with a brief
further discussion of our results.
 
\section{Transfer matrices and fusion hierarchy}\label{sec:transfer}

Sklyanin \cite{Sk} showed for an $N$-site open XXZ spin chain how to
construct a commuting transfer matrix, which here we shall denote by
$t^{(\frac{1}{2},\frac{1}{2})}(u)$, whose auxiliary space as well as each of
its $N$ quantum spaces are spin-$1/2$ (i.e., two-dimensional).  In a
similar way, one can construct a transfer matrix $t^{(j,s)}(u)$ whose
auxiliary space is spin-$j$ ($(2j+1)$-dimensional) and each of its $N$
quantum spaces are spin-$s$ ($(2s+1)$-dimensional), for any $j,s \in 
\{\frac{1}{2},1,\frac{3}{2},\ldots\}$.  The basic
building blocks are so-called fused $R$ and $K^{\mp}$ matrices. 
The former are given by \cite{Ka}, \cite{KRS}-\cite{spinsXXZ} \footnote{Our 
definitions of fused $R$ and $K$ matrices differ from those in 
\cite{YNZ} by certain shifts of the arguments.}
\be
R^{(j,s)}_{\{a\} \{b\}}(u) =
P_{\{a\}}^{+} P_{\{b\}}^{+} 
\prod_{k=1}^{2j}\prod_{l=1}^{2s}
R^{(\frac{1}{2},\frac{1}{2})}_{a_{k} b_{l}}(u+(k+l-j-s-1)\eta)\, 
P_{\{a\}}^{+} P_{\{b\}}^{+} \,,
\label{fusedRmatrix}
\ee 
where $\{a\} = \{a_{1}, \ldots , a_{2j}\}$, $\{b\} = \{b_{1}, \ldots , 
b_{2s}\}$, and $R^{(\frac{1}{2},\frac{1}{2})}(u)$ is given by
\be
R^{(\frac{1}{2},\frac{1}{2})}(u) = \left( \begin{array}{cccc}
	\sh  (u + \eta) &0            &0           &0            \\
	0                 &\sh  u     &\sh \eta  &0            \\
	0                 &\sh \eta   &\sh  u    &0            \\
	0                 &0            &0           &\sh  (u + \eta)
\end{array} \right) \,,
\label{Rmatrix}
\ee 
where $\eta$ is the bulk anisotropy parameter. 
The $R$ matrices in the product (\ref{fusedRmatrix}) are ordered 
in the order of increasing $k$ and $l$.
Moreover, 
$P_{\{a\}}^{+}$ is the symmetric projector
\be
P_{\{a\}}^{+} ={1\over (2j)!} 
\prod_{k=1}^{2j}\left(\sum_{l=1}^{k}{\cal P}_{a_{l}, a_{k}} \right) \,,
\label{projector}
\ee
where ${\cal P}$ is the permutation operator, with ${\cal P}_{a_{k}, 
a_{k}} \equiv 1$; and similarly for $P_{\{b\}}^{+}$. 
For example, for the simple case $(j,s) =(1,\frac{1}{2})$,  
\be
R^{(1,\frac{1}{2})}_{\{ a_{1}, a_{2} \}, b}(u) = 
P_{a_{1}, a_{2}}^{+} 
R^{(\frac{1}{2},\frac{1}{2})}_{a_{1}, b}(u-\frac{1}{2}\eta)\, 
R^{(\frac{1}{2},\frac{1}{2})}_{a_{2}, b}(u+\frac{1}{2}\eta)\,
P_{a_{1}, a_{2}}^{+} \,.
\ee 
The fused $R$ matrices satisfy the Yang-Baxter equations
\be
R^{(j,k)}_{\{a\} \{b\}}(u-v)\, R^{(j,s)}_{\{a\} \{c\}}(u)\, 
R^{(k,s)}_{\{b\} \{c\}}(v) =
R^{(k,s)}_{\{b\} \{c\}}(v)\,  R^{(j,s)}_{\{a\} \{c\}}(u)\, 
R^{(j,k)}_{\{a\} \{b\}}(u-v) \,.
\ee 
We note here for later convenience that the fundamental $R$ matrix 
satisfies the unitarity relation
\be
R^{(\frac{1}{2},\frac{1}{2})}(u) R^{(\frac{1}{2},\frac{1}{2})}(-u) = 
- \xi(u) 1 \,, \qquad \xi(u) = \sh(u+ \eta) \sh(u - \eta) \,.
\label{xi}
\ee

The fused $K^{-}$ matrices are given by \cite{MNR, MN, Zh}
\be
K^{- (j)}_{\{a\}}(u) &=& P_{\{a\}}^{+} \prod_{k=1}^{2j} \Bigg\{ \left[ 
\prod_{l=1}^{k-1} R^{(\frac{1}{2},\frac{1}{2})}_{a_{l}a_{k}}
(2u+(k+l-2j-1)\eta) \right] \non \\
&\times & K^{- (\frac{1}{2})}_{a_{k}}(u+(k-j-\frac{1}{2})\eta) \Bigg\}
P_{\{a\}}^{+} \,,
\label{fusedKmatrix}
\ee 
where $K^{- (\frac{1}{2})}(u)$ is the matrix \cite{dVGR, GZ}
\be
\left( \begin{array}{cc}
2 \left( \sh \alpha_{-} \ch \beta_{-} \ch u +
\ch \alpha_{-} \sh \beta_{-} \sh u \right) &    e^{\theta_{-}} \sh 2u \\
e^{-\theta_{-}} \sh  2u	    &  2 \left( \sh \alpha_{-} \ch \beta_{-} \ch u -
\ch \alpha_{-} \sh \beta_{-} \sh u \right)        
\end{array} \right) \,,
\label{Kminus}
\ee 
where $\alpha_{-}$, $\beta_{-}$ and $\theta_{-}$ are arbitrary 
boundary parameters.
The products of braces $\{ \ldots \}$ in (\ref{fusedKmatrix})
are ordered in the order of increasing $k$. For example, for the case $j=1$,
\be
K^{- (1)}_{a_{1}, a_{2}}(u) &=& P_{a_{1}, a_{2}}^{+} 
K^{- (\frac{1}{2})}_{a_{1}}(u-\frac{1}{2}\eta)\,
R^{(\frac{1}{2},\frac{1}{2})}_{a_{1}, a_{2}}(2u)\,
K^{- (\frac{1}{2})}_{a_{2}}(u+\frac{1}{2}\eta)\,
P_{a_{1}, a_{2}}^{+} \,.
\ee 
The fused $K^{-}$ matrices satisfy the boundary Yang-Baxter equations 
\cite{Ch}
\be
\lefteqn{R^{(j,s)}_{\{a\} \{b\}}(u-v)\, K^{- (j)}_{\{a\}}(u)\,
R^{(j,s)}_{\{a\} \{b\}}(u+v)\, K^{- (j)}_{\{b\}}(u)}\non \\
& & =K^{- (j)}_{\{b\}}(u)\, R^{(j,s)}_{\{a\} \{b\}}(u+v)\,
K^{- (j)}_{\{a\}}(u)\, R^{(j,s)}_{\{a\} \{b\}}(u-v) \,.
\ee

The fused $K^{+}$ matrices are given by
\be
K^{+ (j)}_{\{a\}}(u)  = {1\over f^{(j)}(u)}\,K^{- (j)}_{\{a\}}
(-u-\eta)\Big\vert_{(\alpha_-,\beta_-,\theta_-)\rightarrow
(-\alpha_+,-\beta_+,\theta_+)} \,,
\ee
where the normalization factor
\be
f^{(j)}(u) = \prod_{l=1}^{2j-1}\prod_{k=1}^{l}
[-\xi( 2u + (l+k+1-2j)\eta) ] 
\label{Kplusnormalization}
\ee
is chosen to simplify the form of the fusion hierarchy, see
below (\ref{delta}).

The transfer matrix $t^{(j,s)}(u)$ is given by
\be
t^{(j,s)}(u) = \tr_{\{a\}} K^{+ (j)}_{\{a\}}(u)\,
T^{(j,s)}_{\{a\}}(u)\, K^{- (j)}_{\{a\}}(u)\,
\hat T^{(j,s)}_{\{a\}}(u) \,,
\ee 
where the monodromy matrices are given by products of $N$ fused $R$ 
matrices, 
\be
T^{(j,s)}_{\{a\}}(u) &=& R^{(j,s)}_{\{a\}, \{b^{[N]}\}}(u) \ldots 
R^{(j,s)}_{\{a\}, \{b^{[1]}\}}(u) \,, \non \\
\hat T^{(j,s)}_{\{a\}}(u) &=& R^{(j,s)}_{\{a\}, \{b^{[1]}\}}(u) \ldots
R^{(j,s)}_{\{a\}, \{b^{[N]}\}}(u) \,.
\ee 
These transfer matrices commute for different
values of spectral parameter for any $j , j' \in \{\frac{1}{2}, 1,
\frac{3}{2}, \ldots \}$ and any $s \in \{\frac{1}{2}, 1, \frac{3}{2},
\ldots \}$,
\be
\left[ t^{(j,s)}(u) \,, t^{(j',s)}(u') \right] = 0 \,.
\label{commutativity}
\ee
These transfer matrices also obey the fusion hierarchy  \cite{MNR, 
MN, Zh} \footnote{The derivation of this hierarchy relies on some relations which, to our 
knowledge, have not been proved.  See the Appendix for further details.} 
\be
t^{(j-\frac{1}{2},s)}(u- j\eta)\, t^{(\frac{1}{2},s)}(u) =
t^{(j,s)}(u-(j-\frac{1}{2})\eta)  + \delta^{(s)}(u)\,
t^{(j-1,s)}(u-(j+\frac{1}{2})\eta) \,, 
\label{hierarchy}
\ee
$j = 1,\frac{3}{2},\ldots$, where $t^{(0,s)}=1$, and
$\delta^{(s)}(u)$ is a product of various 
quantum determinants, and is given by
\be
\delta^{(s)}(u) &=& 
2^{4}\left[\prod_{k=0}^{2s-1}\xi(u+(s-k-\frac{1}{2})\eta)\right]^{2N} 
{\sh(2u-2\eta) \sh(2u+2\eta)\over \sh(2u-\eta) \sh(2u+\eta)} \non \\
&\times& \sh(u+\alpha_{-})\sh(u-\alpha_{-})\ch(u+\beta_{-})\ch(u-\beta_{-})\non \\
&\times& \sh(u+\alpha_{+})\sh(u-\alpha_{+})\ch(u+\beta_{+})\ch(u-\beta_{+}) \,.
\label{delta}
\ee
We remark that the normalization factor $f^{(j)}(u)$ (\ref{Kplusnormalization}) 
has been chosen so that the LHS of (\ref{hierarchy}) has coefficient 1.

In the remainder of this section we list some important further
properties of the ``fundamental'' transfer matrix
$t^{(\frac{1}{2},s)}(u)$, which we shall subsequently use to help determine
its eigenvalues. However, it is more convenient to work with a rescaled transfer 
matrix $\tilde t^{(\frac{1}{2},s)}(u)$ defined by
\be
\tilde t^{(\frac{1}{2},s)}(u) = {1\over g^{(\frac{1}{2},s)}(u)^{2N}} 
t^{(\frac{1}{2},s)}(u) \,,
\label{tildet}
\ee
where
\be
g^{(\frac{1}{2},s)}(u) = \prod_{k=1}^{2s-1} \sh(u+(s-k+\frac{1}{2})\eta)
\label{gfunction}
\ee
(which has the crossing symmetry $g^{(\frac{1}{2},s)}(-u -\eta) 
=\pm g^{(\frac{1}{2},s)}(u)$)
is an overall scalar factor of the fused $R$ matrix $R^{(\frac{1}{2},s)}(u)$.
In particular, the rescaled transfer matrix does not vanish at $u=0$
when $s$ is a half-odd integer.

This transfer matrix has the following properties: 
\be
\tilde t^{(\frac{1}{2},s)}(u + i\pi) = \tilde 
t^{(\frac{1}{2},s)}(u) \qquad (i\pi\mbox{ - periodicity}) 
\label{periodicity}
\ee
\be 
\tilde t^{(\frac{1}{2},s)}(-u -\eta) = \tilde 
t^{(\frac{1}{2},s)}(u) \qquad (\mbox{crossing}) 
\label{crossing}
\ee
\be
\tilde t^{(\frac{1}{2},s)}(0) = -2^{3}\sh^{2N}((s+\frac{1}{2})\eta) 
\ch \eta \sh \alpha_{-} \ch \beta_{-} \sh \alpha_{+} \ch \beta_{+} 1
\quad (\mbox{initial condition} )
\label{initial}
\ee
\be
\tilde t^{(\frac{1}{2},s)}(u)\Big\vert_{\eta=0} &=& 
2^{3}\sh^{2N}u \Big[ -\sh \alpha_{-} \ch \beta_{-} \sh \alpha_{+} \ch 
\beta_{+} \ch^{2}u  \non \\
&+& \ch \alpha_{-} \sh \beta_{-} \ch \alpha_{+} \sh 
\beta_{+} \sh^{2}u \non \\
&-& \ch(\theta_{-}-\theta_{+}) \sh^{2}u \ch^{2}u 
\Big] 1 \quad (\mbox{semi-classical limit} )
\label{semiclassical}
\ee
\be 
\tilde t^{(\frac{1}{2},s)}(u) &\sim& -{1\over 2^{2N+1}}e^{(2N+4)u 
+(N+2)\eta} \ch(\theta_{-}-\theta_{+}) 1 \quad \mbox{for} \ u\rightarrow 
+\infty \non \\
& & \qquad \qquad \qquad \qquad \qquad (\mbox{asymptotic behavior} ) 
\label{asymptotic}
\ee
As is well known, due to the commutativity property (\ref{commutativity}), the 
corresponding simultaneous eigenvectors are independent of the spectral 
parameter. Hence, the above properties (\ref{periodicity}) - 
(\ref{asymptotic}) hold also for the corresponding eigenvalues 
$\tilde \Lambda^{(\frac{1}{2},s)}(u)$.

\section{Eigenvalues and Bethe Ansatz equations}\label{sec:BA}

We now proceed to determine the eigenvalues of the fundamental
transfer matrix.  Following \cite{YNZ}, we assume that the limit
\be
\bar Q(u) = \lim_{j \rightarrow \infty} t^{(j-\frac{1}{2},s)}(u- 
j\eta)
\label{Qassumption}
\ee 
exists.  \footnote{For a more rigorous and extensive discussion of $Q$
operators for the six- and eight-vertex models with periodic boundary
conditions, as well numerous earlier references, see \cite{BM}.}
We then immediately obtain from the fusion hierarchy
(\ref{hierarchy}) an equation of the $T-Q$ form for the fundamental
transfer matrix $t^{(\frac{1}{2},s)}(u)$,
\be
\bar Q(u)\, t^{(\frac{1}{2},s)}(u) =
\bar Q(u+\eta)  + \delta^{(s)}(u)\, \bar Q(u-\eta) \,.
\label{TQ0}
\ee
We further assume that the eigenvalues of $\bar Q(u)$ (which we denote by the 
same symbol) have the decomposition $\bar Q(u) = f(u)\, Q(u)$ with
\be
Q(u) = \prod_{j=1}^{M} \sh(u-v_{j}) \sh(u+v_{j}+\eta) \,,
\label{Q}
\ee
which has the crossing symmetry $Q(-u-\eta)=Q(u)$. Here $M$ is some 
nonnegative integer. It follows that 
the eigenvalues $\Lambda^{(\frac{1}{2},s)}(u)$ of $t^{(\frac{1}{2},s)}(u)$
are given by
\be
\Lambda^{(\frac{1}{2},s)}(u) = H_{1}(u) {Q(u+\eta)\over Q(u)} +
H_{2}(u) {Q(u-\eta)\over Q(u)} \,,
\label{TQ1}
\ee
where $H_{1}(u)=f(u+\eta)/f(u)$ and $H_{2}(u)=\delta^{(s)}(u) 
f(u-\eta)/f(u)$; and therefore,
\be
H_{1}(u-\eta)\, H_{2}(u) = \delta^{(s)}(u) \,.
\ee 
The crossing symmetry (\ref{crossing}) together with (\ref{TQ1}) 
imply that 
\be
H_{2}(u) = H_{1}(-u-\eta) \,.
\label{H2eqn}
\ee
We conclude that $H_{1}(u)$ must satisfy
\be
H_{1}(u-\eta)\, H_{1}(-u-\eta) = \delta^{(s)}(u) \,,
\label{H1eqn}
\ee
where $\delta^{(s)}(u)$ is given by (\ref{delta}).

A set of solutions of (\ref{H1eqn}) for $H_{1}(u)$ is given by
\be
& & H_{1}^{(\pm)}(u|\epsilon_{1},\epsilon_{2},\epsilon_{3}) = -2^{2} 
\epsilon_{2} 
\left[\prod_{k=0}^{2s-1}\sh(u+(s-k-\frac{1}{2})\eta)\right]^{2N} 
{\sh(2u)\over \sh(2u+\eta)} \non \\
& & \quad \times
\sh(u \pm \alpha_{-} + \eta)
\ch(u \pm \epsilon_{1}\beta_{-} + \eta)
\sh(u \pm \epsilon_{2}\alpha_{+}+ \eta)
\ch(u \pm \epsilon_{3}\beta_{+} + \eta) \,,
\label{H1}
\ee
where $\epsilon_{1},\epsilon_{2},\epsilon_{3}$ can independently take 
the values $\pm 1$. It follows from (\ref{H2eqn}) that the 
corresponding $H_{2}(u)$ functions are given by 
\be
& & H_{2}^{(\pm)}(u|\epsilon_{1},\epsilon_{2},\epsilon_{3}) = -2^{2} 
\epsilon_{2} 
\left[\prod_{k=0}^{2s-1}\sh(u+(s-k+\frac{1}{2})\eta)\right]^{2N} 
{\sh(2u+2\eta)\over \sh(2u+\eta)} \non \\
& & \quad \times
\sh(u \mp \alpha_{-})
\ch(u \mp \epsilon_{1}\beta_{-})
\sh(u \mp \epsilon_{2}\alpha_{+})
\ch(u \mp \epsilon_{3}\beta_{+}) \,.
\label{H2}
\ee
An argument from \cite{YNZ} (which makes use of the periodicity
(\ref{periodicity})) can again be used to conclude that the
eigenvalues can be uniquely expressed as
\be
\Lambda^{(\frac{1}{2},s)}(u) = 
a H_{1}^{(\pm)}(u|\epsilon_{1},\epsilon_{2},\epsilon_{3}) {Q(u+\eta)\over Q(u)} +
a H_{2}^{(\pm)}(u|\epsilon_{1},\epsilon_{2},\epsilon_{3}) {Q(u-\eta)\over Q(u)} \,,
\label{TQ2}
\ee
up to an overall sign $a = \pm 1$.  Noting that the functions $H_{1}$
(\ref{H1}) and $H_{2}$ (\ref{H2}) have the factor
$g^{(\frac{1}{2},s)}(u)^{2N}$ (\ref{gfunction}) in common, we conclude that the
eigenvalues of $\tilde t^{(\frac{1}{2},s)}(u)$ (\ref{tildet}) are
given by
\be
\tilde \Lambda^{(\frac{1}{2},s)}(u) = 
\tilde H_{1}^{(\pm)}(u|\epsilon_{1},\epsilon_{2},\epsilon_{3}) {Q(u+\eta)\over Q(u)} +
\tilde H_{2}^{(\pm)}(u|\epsilon_{1},\epsilon_{2},\epsilon_{3}) {Q(u-\eta)\over Q(u)} \,,
\label{TQ3}
\ee
where
\be 
\tilde H_{1}^{(\pm)}(u|\epsilon_{1},\epsilon_{2},\epsilon_{3}) &=&  -2^{2} 
\epsilon_{2} 
\sh^{2N} (u-(s-\frac{1}{2})\eta)
{\sh(2u)\over \sh(2u+\eta)} 
\sh(u \pm \alpha_{-} + \eta) \non\\
&\times & 
\ch(u \pm \epsilon_{1}\beta_{-} + \eta)
\sh(u \pm \epsilon_{2}\alpha_{+}+ \eta)
\ch(u \pm \epsilon_{3}\beta_{+} + \eta) \,,   \label{tildeH1}\\
\tilde H_{2}^{(\pm)}(u|\epsilon_{1},\epsilon_{2},\epsilon_{3}) &=& -2^{2} 
\epsilon_{2} 
\sh^{2N} (u+(s+\frac{1}{2})\eta)
{\sh(2u+2\eta)\over \sh(2u+\eta)}
\sh(u \mp \alpha_{-}) \non\\
&\times & 
\ch(u \mp \epsilon_{1}\beta_{-})  
\sh(u \mp \epsilon_{2}\alpha_{+})
\ch(u \mp \epsilon_{3}\beta_{+}) \,.\label{tildeH2}
\ee
We have fixed the overall sign $a=+1$ in (\ref{TQ3}) using the initial 
condition (\ref{initial}).

The expression (\ref{TQ3}) for $\tilde \Lambda^{(\frac{1}{2},s)}(u)$ 
is also consistent with the asymptotic behavior (\ref{asymptotic}) if the 
boundary parameters satisfy the constraint
\be
\alpha_{-}+\epsilon_{1}\beta_{-}+\epsilon_{2}\alpha_{+}+\epsilon_{3}\beta_{+} 
= \epsilon_{0}(\theta_{-}-\theta_{+}) + \eta\, k + 
\frac{1}{2}(1-\epsilon_{2}) i \pi \quad \mbox{mod} (2i\pi) \,,
\label{constraint}
\ee 
where also $\epsilon_{0} = \pm 1$; and if $M$ (appearing in the 
expression (\ref{Q}) for $Q(u)$) is given by
\be
M = s N - \frac{1}{2} \mp \frac{k}{2}
\label{M} \,.
\ee
The requirement that $M$ be an integer evidently implies that
\be
s N - \frac{1}{2} \mp \frac{k}{2} = \mbox{integer}\,.
\label{kconstraint}
\ee
In particular, for $s$ an integer, $k$ is an odd integer; and for $s$ a
half-odd integer, $k$ is odd (even) if $N$ is even (odd), respectively.

Finally, the expression (\ref{TQ3}) for $\tilde
\Lambda^{(\frac{1}{2},s)}(u)$ is also consistent with the
semi-classical limit (\ref{semiclassical}) if the $\{ \epsilon_{i}\}$ satisfy
the constraint
\be
\epsilon_{1}\, \epsilon_{2}\, \epsilon_{3} = +1 \,.
\label{epsilonconstraint}
\ee

To summarize: if the boundary parameters ($\alpha_{\pm}, \beta_{\pm},
\theta_{\pm}$) satisfy the constraints (\ref{constraint}),
(\ref{kconstraint}) and (\ref{epsilonconstraint}) for some choice
($\pm 1$) of $\{ \epsilon_{i}\}$ and for some appropriate value of
$k$, then the eigenvalues of $\tilde t^{(\frac{1}{2},s)}(u)$
(\ref{tildet}) are given by
\be
\tilde \Lambda^{(\frac{1}{2},s) (\pm)}(u) = 
\tilde H_{1}^{(\pm)}(u|\epsilon_{1},\epsilon_{2},\epsilon_{3}) 
{Q^{(\pm)}(u+\eta)\over Q^{(\pm)}(u)} +
\tilde H_{2}^{(\pm)}(u|\epsilon_{1},\epsilon_{2},\epsilon_{3}) 
{Q^{(\pm)}(u-\eta)\over Q^{(\pm)}(u)} \,,
\label{TQfinal}
\ee
where $\tilde H_{1}^{(\pm)}(u|\epsilon_{1},\epsilon_{2},\epsilon_{3})$ and 
$\tilde H_{2}^{(\pm)}(u|\epsilon_{1},\epsilon_{2},\epsilon_{3})$ are 
given by (\ref{tildeH1}), (\ref{tildeH2}), and 
\be
Q^{(\pm)}(u) = \prod_{j=1}^{M^{(\pm)}} \sh(u-v^{(\pm)}_{j}) 
\sh(u+v^{(\pm)}_{j}+\eta) \,, \quad M^{(\pm)} = s N - \frac{1}{2} \mp 
\frac{k}{2}\,.
\ee
The parameters $\{ v^{(\pm)}_{j} \}$ are roots of the corresponding 
Bethe Ansatz equations,
\be
{\tilde H_{2}^{(\pm)}(v^{(\pm)}_{j}|\epsilon_{1},\epsilon_{2},\epsilon_{3})
\over
\tilde H_{2}^{(\pm)}(-v^{(\pm)}_{j}-\eta|\epsilon_{1},\epsilon_{2},\epsilon_{3})}
=-{Q^{(\pm)}(v^{(\pm)}_{j}+\eta)\over Q^{(\pm)}(v^{(\pm)}_{j}-\eta)} 
\,, \qquad j = 1, \ldots, M^{(\pm)} \,;
\ee
or, more explicitly, 
\be
\lefteqn{\left(
{\sh(\tilde  v^{(\pm)}_{j} + s \eta) \over 
\sh (\tilde  v^{(\pm)}_{j} - s \eta)}\right)^{2N}
{\sh(2\tilde v^{(\pm)}_{j} + \eta) \over 
\sh (2\tilde  v^{(\pm)}_{j} - \eta)}
{\sh(\tilde  v^{(\pm)}_{j} \mp \alpha_{-} - {\eta\over 2}) \over 
\sh (\tilde  v^{(\pm)}_{j} \pm \alpha_{-} + {\eta\over 2})}
{\ch(\tilde  v^{(\pm)}_{j} \mp \epsilon_{1} \beta_{-} - {\eta\over 2}) \over 
\ch (\tilde  v^{(\pm)}_{j} \pm \epsilon_{1} \beta_{-} + {\eta\over 2})}} \non \\
& & \times 
{\sh(\tilde  v^{(\pm)}_{j} \mp \epsilon_{2}\alpha_{+} - {\eta\over 2}) \over 
\sh (\tilde  v^{(\pm)}_{j} \pm \epsilon_{2}\alpha_{+} + {\eta\over 2})}
{\ch(\tilde  v^{(\pm)}_{j} \mp \epsilon_{3}\beta_{+} - {\eta\over 2}) \over 
\ch (\tilde  v^{(\pm)}_{j} \pm \epsilon_{3}\beta_{+} + {\eta\over 2})} \non \\
&=& - \prod_{k=1}^{M^{(\pm)}} 
{\sh(\tilde  v^{(\pm)}_{j} -  \tilde v^{(\pm)}_{k} + \eta) \over 
\sh (\tilde  v^{(\pm)}_{j} -  \tilde v^{(\pm)}_{k} - \eta)}
{\sh(\tilde  v^{(\pm)}_{j} +  \tilde v^{(\pm)}_{k} + \eta) \over 
\sh (\tilde  v^{(\pm)}_{j} +  \tilde v^{(\pm)}_{k} - \eta)} \,, 
\quad j = 1 \,, \cdots \,, M^{(\pm)} \,, 
\label{BAEs}
\ee
where $\tilde  v^{(\pm)}_{j}  \equiv v^{(\pm)}_{j} +\eta/2$.

We have investigated the completeness of this solution for small
values of $s$ and $N$ numerically using a method developed by McCoy
and his collaborators (see, e.g., \cite{Mc}), and further explained in
\cite{NR}.  We find that, for $k \ge 2sN+1$ (and therefore $M^{(+)}\le
-1$), all the eigenvalues of $\tilde t^{(\frac{1}{2},s)}(u)$
are given by $\tilde \Lambda^{(\frac{1}{2},s)
(-)}(u)$.  Similarly, for $k \le -(2sN+1)$ (and therefore $M^{(-)}\le
-1$), all the eigenvalues are given by $\tilde \Lambda^{(\frac{1}{2},s)
(+)}(u)$.  Moreover, for $|k| \le 2sN-1$ (and therefore both $M^{(+)}$
and $M^{(-)}$ are nonnegative), both $\tilde
\Lambda^{(\frac{1}{2},s) (-)}(u)$ and $\tilde \Lambda^{(\frac{1}{2},s)
(+)}(u)$ are needed to obtain all the eigenvalues.  \footnote{For the 
conventional situation that the boundary parameters ($\alpha_{\pm}, 
\beta_{\pm}, \theta_{\pm}$) are finite and independent of $N$, the 
constraint (\ref{constraint}) requires that also $k$ be finite and independent 
of $N$, in which case $|k| \le 2sN-1$ for $N \rightarrow \infty$.}

Some sample results are summarized in Tables
\ref{table:n2s1}-\ref{table:n3s32}.  For example, let us consider
Table \ref{table:n2s1}, which is for the case $N=2$, $s=1$.  According
to (\ref{kconstraint}), $k$ must be odd for this case.  For such
values of $k$, the number of eigenvalues of $\tilde
t^{(\frac{1}{2},s)}(u)$ which we found that are given by $\tilde
\Lambda^{(\frac{1}{2},s) (-)}(u)$ and $\tilde \Lambda^{(\frac{1}{2},s)
(+)}(u)$ in (\ref{TQfinal}) are listed in the second and third
columns, respectively.  Notice that, for each row of the table, the
sum of these two entries is $9$, which coincides with the total number
of eigenvalues ($3^{2}$).  A similar result can be readily seen in the
other tables.  \footnote{It would be interesting to find a formula
that would generate the entries in these tables for general values of
$N, s$ and $k$.} These results strongly support the conjecture that
$\tilde \Lambda^{(\frac{1}{2},s) (-)}(u)$ and $\tilde
\Lambda^{(\frac{1}{2},s) (+)}(u)$ in (\ref{TQfinal}) together give the
complete set of $(2s+1)^{N}$ eigenvalues of the transfer matrix
$\tilde t^{(\frac{1}{2},s)}(u)$ for $|k| \le 2sN-1$.

It may be worth noting that for $|k|=2sN-1$ (and, therefore, either
$M^{(+)}$ or $M^{(-)}$ vanishes), the Bethe Ansatz with vanishing $M$
still gives 1 eigenvalue.  (See again Tables
\ref{table:n2s1}-\ref{table:n3s32}.)  In the algebraic Bethe Ansatz
approach, the corresponding eigenvector would presumably be a ``bare''
reference state.

We observe that for $s=1/2$, our solution coincides with the one in \cite{YNZ}.
Our solution is similar to the one given by Doikou \cite{Do2, Do3},
but with a more general constraint on the boundary parameters, and with
a second set of BAEs which is necessary for completeness.

\section{Hamiltonian for $s=1$}\label{sec:Ham}

In this section we give an explicit expression for the spin-1
Hamiltonian, its relation to the transfer matrix, and its
eigenvalues in terms of Bethe roots.  In order to construct the
integrable Hamiltonian for the case $s=1$, we need the transfer matrix
with spin-1 in {\it both} auxiliary and quantum spaces, i.e.,
$t^{(1,1)}(u)$.  Since we seek an explicit expression for the
Hamiltonian in terms of spin-1 generators of $su(2)$ (which are
$3\times 3$ matrices), we now perform similarity transformations on
the fused $R$ and $K$ matrices which bring them to row-reduced form,
and then remove all null rows and columns.  That is,
\be
R^{(1,1)}_{reduced}(u) &=&
A_{a_{1}, a_{2}} A_{b_{1}, b_{2}}\, 
R^{(1,1)}_{\{ a_{1}, a_{2} \}, \{ b_{1}, b_{2} \}}(u)\,
A^{-1}_{a_{1}, a_{2}} A^{-1}_{b_{1}, b_{2}}\,, \non \\
K^{\mp (1)}_{reduced}(u)  &=&
A_{a_{1}, a_{2}}\, 
K^{\mp (1)}_{a_{1}, a_{2}}(u)\,
A^{-1}_{a_{1}, a_{2}} \,,
\ee
where
\be
A=\left( \begin{array}{rrrr}
	1    &0            &0            &0            \\
	0    &\frac{1}{2}  &\frac{1}{2}  &0            \\
	0    &\frac{1}{2}  &-\frac{1}{2}  &0           \\
	0                 &0            &0           &1
\end{array} \right) \,.
\ee 
The reduced $R$ and $K$ matrices are $9 \times 9$ 
(instead of $16 \times 16$) and $3 \times 3$ (instead of $4 \times 4$) matrices, respectively.
It is convenient to make a further similarity (``gauge'') 
transformation which brings these matrices to a more symmetric form,
\be
R^{(1,1)\ gt}_{reduced}(u) &=& (B \otimes B) \, R^{(1,1)}_{reduced}(u)\, 
(B^{-1} \otimes B^{-1}) \non \\
K^{\mp (1)\ gt}_{reduced}(u)  &=&  B \, K^{\mp (1)}_{reduced}(u)\, B^{-1}
\,,
\ee
where $B$ is the $3 \times 3$  diagonal matrix
\be
B = \mbox{diag} (1\,, -\sqrt{2 \ch \eta} \,, 1) \,.
\ee
We define $t^{(1,1)\ gt}(u)$ to be the transfer matrix constructed 
with these $R$ and $K$ matrices, i.e.,
\be
t^{(1,1)\ gt}(u) = \tr_{0} K^{+ (1)\ gt}_{reduced\ 0}(u)\,
T^{(1,1)\ gt}_{reduced\ 0}(u)\, K^{- (1)\ gt}_{reduced\ 0}(u)\,
\hat T^{(1,1)\ gt}_{reduced\ 0}(u) \,,
\ee 
where the monodromy matrices are constructed as usual from the $R^{(1,1)\ 
gt}_{reduced}(u)$'s, and the auxiliary space is now denoted by ``0''. 
Since $t^{(1,1)}(u)$ and $t^{(1,1)\ gt}(u)$ are related by a similarity 
transformation, they have the same eigenvalues. Finally, it is 
again more convenient to work with a rescaled transfer matrix,
\be
\tilde t^{(1,1)\ gt}(u) = {\sh(2u) \sh(2u+2\eta)\over [\sh u 
\sh(u+\eta)]^{2N}}\, 
t^{(1,1)\ gt}(u) \,,
\label{rescaled}
\ee
which, in particular, does not vanish at $u=0$.

As noted by Sklyanin \cite{Sk}, the Hamiltonian $H$ is proportional to
the first derivative of the transfer matrix
\be
H = c_{1} {d\over du}\tilde t^{(1,1)\ gt}(u) \Big\vert_{u=0} + c_{0}1
\label{Htrelation}
\,,
\ee
where
\be 
c_{1}&=&-\ch \eta \Big\{ 16 [\sh 2\eta \sh \eta]^{2N} \sh 3\eta 
\sh(\alpha_{-}-{\eta\over 2})\sh(\alpha_{-}+{\eta\over 2})
\ch(\beta_{-}-{\eta\over 2})\ch(\beta_{-}+{\eta\over 2})\non \\
&\times& \sh(\alpha_{+}-{\eta\over 2})\sh(\alpha_{+}+{\eta\over 2})
\ch(\beta_{+}-{\eta\over 2})\ch(\beta_{+}+{\eta\over 2})\Big\}^{-1}
\label{c1}
\,.
\ee
We choose $c_{1}$ so that the bulk terms of the Hamiltonian have a
conventional normalization (see (\ref{bulkhamiltonian})), and we choose
$c_{0}$ so that there is no additive constant term in the expression
(\ref{Hamiltonian}) for the Hamiltonian,
\be
H = \sum_{n=1}^{N-1}H_{n,n+1} + H_{b} \,.
\label{Hamiltonian}
\ee
The bulk terms $H_{n,n+1}$ are given by \cite{ZF}
\be 
H_{n,n+1} &=&  \sigma_{n} - (\sigma_{n})^{2}
+ 2 \sh^2 \eta \left[ \sigma_{n}^{z} + (S^z_n)^2
+ (S^z_{n+1})^2 - (\sigma_{n}^{z})^2 \right] \non \\
&-& 4 \sh^2 (\frac{\eta}{2})  \left( \sigma_{n}^{\bot} \sigma_{n}^{z}
+ \sigma_{n}^{z} \sigma_{n}^{\bot} \right) \,, \label{bulkhamiltonian}
\ee 
where
\be
\sigma_{n} = \vec S_n \cdot \vec S_{n+1} \,, \quad
\sigma_{n}^{\bot} = S^x_n S^x_{n+1} + S^y_n S^y_{n+1}  \,, \quad
\sigma_{n}^{z} = S^z_n S^z_{n+1} \,, 
\ee 
and $\vec S$ are the standard spin-1 generators of $su(2)$. The 
boundary terms $H_{b}$ have the form \cite{IOZ}
\be 
H_{b} &=& a_{1} (S^{z}_{1})^{2}  + a_{2} S^{z}_{1} 
+  a_{3} (S^{+}_{1})^{2}  +  a_{4} (S^{-}_{1})^{2}  +
a_{5} S^{+}_{1}\, S^{z}_{1}  + a_{6}  S^{z}_{1}\, S^{-}_{1} \non \\
&+& a_{7}  S^{z}_{1}\, S^{+}_{1} + a_{8} S^{-}_{1}\, S^{z}_{1} 
+ (a_{j} \leftrightarrow b_{j} \mbox{ and } 1 \leftrightarrow N) \,,
\ee
where $S^{\pm} = S^{x} \pm i S^{y}$. The coefficients $\{ a_{i} \}$ 
of the boundary terms at site 1 are given in 
terms of the boundary parameters ($\alpha_{-}, \beta_{-},
\theta_{-}$) by
\be
a_{1} &=& \frac{1}{4} a_{0} \left(\ch 2\alpha_{-} - \ch 
2\beta_{-}+\ch \eta \right) \sh 2\eta 
\sh \eta \,,\non \\
a_{2} &=& \frac{1}{4} a_{0} \sh 2\alpha_{-} \sh 2\beta_{-} \sh 2\eta \,, \non \\
a_{3} &=& -\frac{1}{8} a_{0} e^{2\theta_{-}} \sh 2\eta 
\sh \eta \,, \non \\
a_{4} &=& -\frac{1}{8} a_{0} e^{-2\theta_{-}} \sh 2\eta 
\sh \eta \,, \non \\
a_{5} &=&  a_{0} e^{\theta_{-}} \left(
\ch \beta_{-}\sh \alpha_{-} \ch {\eta\over 2} +
\ch \alpha_{-}\sh \beta_{-} \sh {\eta\over 2} \right)
\sh \eta \ch^{\frac{3}{2}}\eta \,, \non \\
a_{6} &=&  a_{0} e^{-\theta_{-}} \left(
\ch \beta_{-}\sh \alpha_{-} \ch {\eta\over 2} +
\ch \alpha_{-}\sh \beta_{-} \sh {\eta\over 2} \right)
\sh \eta \ch^{\frac{3}{2}}\eta \,, \non \\
a_{7} &=&  -a_{0} e^{\theta_{-}} \left(
\ch \beta_{-}\sh \alpha_{-} \ch {\eta\over 2} -
\ch \alpha_{-}\sh \beta_{-} \sh {\eta\over 2} \right)
\sh \eta \ch^{\frac{3}{2}}\eta \,, \non \\
a_{8} &=&  -a_{0} e^{-\theta_{-}} \left(
\ch \beta_{-}\sh \alpha_{-} \ch {\eta\over 2} -
\ch \alpha_{-}\sh \beta_{-} \sh {\eta\over 2} \right)
\sh \eta \ch^{\frac{3}{2}}\eta \,,
\ee
where 
\be
a_{0}= \left[
\sh(\alpha_{-}-{\eta\over 2})\sh(\alpha_{-}+{\eta\over 2})
\ch(\beta_{-}-{\eta\over 2})\ch(\beta_{-}+{\eta\over 2})\right]^{-1} 
\,.
\ee
Moreover, the coefficients $\{ b_{i} \}$ of the boundary terms at 
site $N$ are given in 
terms of the boundary parameters ($\alpha_{+}, \beta_{+}, \theta_{+}$) by
\be
b_{i} = a_{i}\Big\vert_{\alpha_{-}\rightarrow \alpha_{+}, 
\beta_{-}\rightarrow -\beta_{+}, \theta_{-}\rightarrow \theta_{+}} \,.
\ee

We now proceed to find an expression for the energies in terms of the
Bethe roots.  It follows from (\ref{Htrelation}) that the eigenvalues
of the Hamiltonian are given by
\be
E = c_{1} {d\over du}\tilde \Lambda^{(1,1)\ gt}(u) \Big\vert_{u=0} + c_{0} 
\,.
\ee
Furthermore,
\be
\tilde \Lambda^{(1,1)\ gt}(u) = {\sh(2u) \sh(2u+2\eta)\over [\sh u 
\sh(u+\eta)]^{2N}}\, 
\Lambda^{(1,1)}(u) \,,
\ee
where we have used (\ref{rescaled}) and the fact that 
$\Lambda^{(1,1)\ gt}(u) = \Lambda^{(1,1)}(u)$.
>From the fusion hierarchy (\ref{hierarchy}) with $j=s=1$, 
we obtain (after performing the shift $u \rightarrow u  + \eta/2$) 
the following relation between $\Lambda^{(1,1)}(u)$ and 
$\Lambda^{(\frac{1}{2},1)}(u)$:
\be
\Lambda^{(1,1)}(u) = \Lambda^{(\frac{1}{2},1)}(u-{\eta\over 2})\,
\Lambda^{(\frac{1}{2},1)}(u+{\eta\over 2}) - 
\delta^{(1)}(u+{\eta\over 2}) \,.
\ee
Recalling (\ref{tildet}) 
\be
\Lambda^{(\frac{1}{2},1)}(u) = g^{(\frac{1}{2},1)}(u)^{2N}\, 
\tilde\Lambda^{(\frac{1}{2},1)}(u) \,,
\ee
and also our result (\ref{TQfinal}) for
$\tilde\Lambda^{(\frac{1}{2},1)}(u)$, we finally arrive at an
expression for the energies in terms of the Bethe roots
\be
E= \sh^{2}(2\eta) \sum_{j=1}^{M^{(\pm)}}
{1\over \sh(\tilde v_{j}^{(\pm)} - \eta)\, \sh(\tilde v_{j}^{(\pm)} + 
\eta)} + N \left({\sh 3\eta\over \sh \eta} -3\right) + c^{(\pm)} \,,
\label{energies}
\ee
where $\tilde v^{(\pm)}_{j} \equiv v^{(\pm)}_{j} +\eta/2$ as in
(\ref{BAEs}), and $c^{(\pm)}$ are constants whose cumbersome
expressions we refrain from presenting here.  (These constants
are independent of $N$, but do depend on the bulk and boundary
parameters and on $\{ \epsilon_{i} \}$.)

We have verified that the energies given by the Bethe Ansatz
(\ref{energies}) coincide with those obtained by direct
diagonalization of the Hamiltonian (\ref{Hamiltonian}) for values of
$N$ up to 4.  For the case $N=3$, some sample results are presented in
Tables \ref{table:energiesM}, \ref{table:energiesP}, for boundary
parameter values corresponding to $k=1$.  Hence, as already noted in
Table \ref{table:n3s1}, 17 levels are obtained with $\tilde
\Lambda^{(\frac{1}{2},1) (-)}(u)$ and are listed in Table
\ref{table:energiesM}; and 10 levels are obtained with $\tilde
\Lambda^{(\frac{1}{2},1) (+)}(u)$ and are listed in Table
\ref{table:energiesP}.  Together, they give all 27 energies obtained
by direct diagonalization of the Hamiltonian.

For $s>1$, it is in principle possible to proceed in a similar way.
However, the computations become significantly more cumbersome, and we
shall not pursue them further.

\section{Discussion}\label{sec:conclude}

We have found a Bethe Ansatz solution of the open spin-$s$ XXZ chain
with general integrable boundary terms (\ref{TQfinal}) - (\ref{BAEs}),
which is valid for generic values of the bulk anisotropy parameter
$\eta$, provided that the boundary parameters satisfy the constraints
(\ref{constraint}), (\ref{kconstraint}) and (\ref{epsilonconstraint}).
We have presented numerical evidence that this solution is complete,
and we have explicitly exhibited the Hamiltonian and its relation to
the transfer matrix for the case $s=1$.  The apparent correctness of
this solution provides further support for the validity of the
identification (\ref{Qassumption}) of the $Q$ operator as a transfer
matrix whose auxiliary-space spin tends to infinity.

A drawback of this solution is that for $|k| \le 2sN-1$, one does not know
a priori in which of the two ``sectors'' (i.e., $\tilde \Lambda^{(\frac{1}{2},s)
(-)}(u)$ or $\tilde \Lambda^{(\frac{1}{2},s) (+)}(u)$) a given level
-- such as the ground state -- will be.  For the case $s=1/2$,
alternative Bethe Ansatz-type solutions have been found \cite{MNS}
which do not suffer from this difficulty, and for which the boundary
parameters do not need to obey the constraints (\ref{constraint}),
(\ref{kconstraint}) and (\ref{epsilonconstraint}).  However, those
solutions hold only for values of bulk anisotropy corresponding to
roots of unity.  (For yet another approach to this problem, see
\cite{BK}.)  Perhaps such solutions can also be generalized to higher
values of $s$.

Part of our motivation for considering this problem comes from the
relation of the $s=1$ case to the supersymmetric sine-Gordon (SSG)
model \cite{SSG}, in particular, its boundary version \cite{BSSG}.
Indeed, Bethe Ansatz solutions of the spin-1 XXZ chain have been used
to derive nonlinear integral equations (NLIEs) for the SSG model on a
circle \cite{NLIE} and on an interval with Dirichlet boundary
conditions \cite{ANS}.  With our solution in hand, one can now try to
derive an NLIE for the SSG model on an interval with general
integrable boundary conditions, and try to make contact with
previously proposed boundary actions and boundary $S$ matrices
\cite{BSSG}.

\section*{Acknowledgments}
RN gratefully acknowledges the hospitality and support extended to 
him at LAPTH, and discussions with V. Bazhanov and M. Wachs.
This work was supported in part by the
National Science Foundation under Grants PHY-0244261 and PHY-0554821.

\begin{appendix}

\section{Conjectured relations for the fusion hierarchy}
 
One way to derive the fusion hierarchy (\ref{hierarchy}) relies on the 
existence of the relations
\be
P^{+}_{12\ldots 2j-1}\, P^{-}_{12\ldots 2j}\, P^{+}_{12 \ldots 2j-1}
=P^{+}_{12 \ldots 2j-2} P^{-}_{2j-1, 2j} 
+ \sum_{k=1}^{2j-1} P^{+}_{k, k+1}\, X^{(k)}\, P^{-}_{k, k+1} \,,
\ j=\frac{3}{2},2, \ldots\ 
\label{conjecture}
\ee
where $ P^{-}_{12\ldots n} \equiv 1 -  P^{+}_{12\ldots n}$, 
for some set of matrices $\{ X^{(1)}\,, \ldots\,,  X^{(2j-1)}\}$.
To our knowledge, these relations have not been proved for general 
values of $j$. However, we have verified them up to $j=3$. In 
particular, for $j=3/2$,
\be
X^{(1)}=\frac{1}{3}\Big( {\cal P}_{23}-{\cal P}_{13}\Big) \,, \qquad
X^{(2)}=\frac{1}{3}\Big( {\cal P}_{12}-{\cal P}_{13}\Big) \,.
\ee 
This result is equivalent to an identity found by Kulish and Sklyanin,
see Eq.  (4.15) in the second reference of \cite{KRS}.  For $j=2$, we 
find
\be
X^{(1)}&=& \frac{1}{12} \big( {\cal P}_{23} - {\cal P}_{13} + {\cal P}_{24}
- {\cal P}_{14} + 2 {\cal P}_{13} {\cal P}_{24} - 2 {\cal P}_{14} {\cal
P}_{23} \big) \,, \non \\
X^{(2)}&=& \frac{1}{6} \big( {\cal P}_{13} - {\cal P}_{12} + {\cal P}_{34}
- {\cal P}_{24} + 2{\cal P}_{12}{\cal P}_{34} - 2{\cal P}_{13} {\cal
P}_{24} \big) \,, \non \\
X^{(3)}&=& \frac{1}{4} \big( {\cal P}_{13} - {\cal P}_{14} + {\cal P}_{23}
- {\cal P}_{24} \big) \,.
\ee
Moreover, for $j=5/2$,
\be
X^{(1)}&=&
  \frac{3}{5} {\cal P}_{15} {\cal P}_{14}
+ {\cal P}_{15} {\cal P}_{24}    
- \frac{2}{5}{\cal P}_{15} {\cal P}_{34}  
+ \frac{1}{2} {\cal P}_{35} {\cal P}_{14}  \non \\
&-& \frac{1}{10}{\cal P}_{35} {\cal P}_{13}  
- \frac{9}{10}{\cal P}_{45} {\cal P}_{14} 
- \frac{1}{6} {\cal P}_{45} {\cal P}_{13}  
+\frac{4}{15} {\cal P}_{14} {\cal P}_{13}  \,,
\non \\
X^{(2)}&=& 
 \frac{1}{5} {\cal P}_{15} {\cal P}_{24} 
+\frac{1}{15}{\cal P}_{15}  {\cal P}_{13} 
-\frac{2}{5}{\cal P}_{25} {\cal P}_{14} 
+\frac{2}{5}{\cal P}_{25} {\cal P}_{24}
+ \frac{4}{15}{\cal P}_{25} {\cal P}_{34} \non \\
&-& \frac{2}{15}{\cal P}_{25}  {\cal P}_{13}
- \frac{7}{15}{\cal P}_{45} {\cal P}_{24}  
-\frac{1}{15}{\cal P}_{45} {\cal P}_{13}  
+\frac{2}{15}{\cal P}_{14}  {\cal P}_{13} \,,  \\
X^{(3)}&=& 
 \frac{3}{10} {\cal P}_{15} {\cal P}_{14}  
-\frac{3}{10} {\cal P}_{15} {\cal P}_{24} 
+ \frac{1}{10} {\cal P}_{25} {\cal P}_{14}  
+\frac{3}{10} {\cal P}_{25} {\cal P}_{24} 
\non \\
&-&  \frac{2}{5} {\cal P}_{35} {\cal P}_{14} 
-\frac{1}{5} {\cal P}_{35} {\cal P}_{34}  
- \frac{2}{5} {\cal P}_{35} {\cal P}_{12} \,, \non \\
X^{(4)}&=&
-\frac{4}{5} {\cal P}_{15} {\cal P}_{14}  
+ \frac{2}{5} {\cal P}_{15} {\cal P}_{24} 
+ \frac{2}{5} {\cal P}_{15} {\cal P}_{34}  
-\frac{2}{5} {\cal P}_{15} {\cal P}_{23}  \,. \non
\ee 
We omit our lengthy results for $j=3$.
We emphasize that the matrices $\{ X^{(k)} \}$ are by no means 
unique: for instance, only the antisymmetric part 
$X^{(k)}_{1\ldots k,k+1\ldots 2j}-X^{(k)}_{1\ldots k+1,k\ldots 2j}$ 
matters in the expression of $X^{(k)}$.
In fact, for the purpose of deriving the fusion hierarchy, the 
explicit matrices are not important, as the terms $\sum_{k} P^{+}_{k, 
k+1}\, X^{(k)}\, P^{-}_{k, k+1}$ in (\ref{conjecture}) do not contribute.
What is important is only the fact that such matrices exist.

We also observe that the symmetrizer $P^+_{1\ldots n}$ (\ref{projector})
can be expressed as a sum of products of
\textsl{commuting}
permutations:
\begin{equation}
P^{+}_{1\ldots n} = a^{(n)}_{0} +\sum_{\ell=1}^{[n/2]} a^{(n)}_{\ell}
\sum_{\sigma\in\cS_{n}} \cP_{\sigma(1)\,\sigma(2)}\,\cP_{\sigma(3)\,\sigma(4)}\cdots 
\cP_{\sigma(2\ell-1)\,\sigma(2\ell)}
\label{fund-exp}
\end{equation}
where $\cS_{n}$ is the permutation group of $n$ indices.
The coefficients $a^{(n)}_{\ell}$ are given by the recursion 
relations
\begin{eqnarray}
a^{(n+1)}_{0} &=& \frac{1}{n+1}\left(a^{(n)}_{0} -n!\, 
a^{(n)}_{1}\right) \label{rec-a0}\\
a^{(n+1)}_{1} &=& \frac{1}{(n+1)^2}\left(\frac{1}{(n-1)!}\, a^{(n)}_{0} 
+4\, a^{(n)}_{1} -2\, a^{(n)}_{2}\right) \\
a^{(n+1)}_{\ell} &=& \frac{3\ell+1}{(n+1)^2}\, a^{(n)}_{\ell}
 -\frac{\ell+1}{(n+1)^2}\, a^{(n)}_{\ell+1} 
+\frac{n+2-2\ell}{(n+1)^2}\, a^{(n)}_{\ell-1} \,,\quad 
2\leq \ell\leq\left[\frac{n}{2}\right]\qquad\\
a^{(n+1)}_{[n/2]+1} &=& \frac{n-2[{n}/{2}]}{(n+1)^2}\, 
a^{(n)}_{[n/2]} \label{rec-afin}
\end{eqnarray}
with $a^{(1)}_{0}=1$ and the convention $a^{(n)}_{\ell}=0$ when
$\ell>[n/2]$. 
Above, $[\ ]$ denotes the integer part. 
Note that $a^{(1)}_{0}=1$ is consistent with our previous convention
$\cP_{11}=1$, which implies $P^+_{1}=1$ and $P^-_{1}=0$.
From the recursion, one then finds e.g. $a^{(2)}_{0}=\half$ and 
$a^{(2)}_{1}=\frac{1}{4}$, which reproduces the result
$P^+_{12}=\half(1+\cP_{12})$. 
Let us stress that, because of the 
sum on all permutations in $\cS_{n}$, a term containing $\ell$
permutations has a multiplicity 
$\ell!\,(n-2\ell)!\,2^\ell$ with respect to a 
`reduced' expression, where all the terms appear just once.
Identity (\ref{fund-exp}) may help in the proof of the conjecture
(\ref{conjecture}).

One can prove relation (\ref{fund-exp}) by recursion. Starting from the relation 
at 
level $n$, the recursion relation
$$
(n+1)\,P^{+}_{1\ldots n+1} = 
\left(1+\sum_{\ell=1}^{n}\cP_{\ell,n+1}\right)\, P^{+}_{1\ldots n} 
 = 
P^{+}_{1\ldots n}\,\left(1+\sum_{\ell=1}^{n}\cP_{\ell,n+1}\right) 
$$
and the identity (valid for any spaces $a$, $b$, $c$)
$$
\cP_{ab}\Big(\cP_{ab}+\cP_{ac}+\cP_{bc}\Big) = \cP_{ab}+\cP_{ac}+\cP_{bc}
$$
show that the relation is also obeyed at level $n+1$. A careful 
analysis of the different terms leads to the relations 
(\ref{rec-a0}-\ref{rec-afin}).
\end{appendix}

\newpage

\begin{table}[htb] 
  \centering
  \begin{tabular}{|c|c|c|}\hline
   k & no. given by $\tilde \Lambda^{(\frac{1}{2},1) 
   (-)}(u)$ & no. given by 
   $\tilde \Lambda^{(\frac{1}{2},1) (+)}(u)$\\
    \hline
      5 & 9 & 0\\
      3 & 8 & 1\\
      1 & 6 & 3\\
      -1 & 3 & 6\\
      -3 & 1 & 8\\
      -5 & 0 & 9\\
     \hline
	\end{tabular}
	\caption[xxx]{\parbox[t]{0.8\textwidth}{
	Number of eigenvalues of $\tilde t^{(\frac{1}{2},s)}(u)$ 
	given by $\tilde \Lambda^{(\frac{1}{2},s) (\pm)}(u)$ for
	$N=2$, $s=1$. The total number of eigenvalues is $(2s+1)^{N}=3^{2}=9$.}
	}
       \label{table:n2s1}
     \end{table}
\begin{table}[htb] 
  \centering
  \begin{tabular}{|c|c|c|}\hline
   k & no. given by $\tilde \Lambda^{(\frac{1}{2},1) 
   (-)}(u)$ & no. given by $\tilde \Lambda^{(\frac{1}{2},1) 
   (+)}(u)$\\
    \hline
      7 & 27 & 0\\
      5 & 26 & 1 \\
      3 & 23 & 4\\
      1 & 17 & 10\\
      -1 & 10 & 17\\
      -3 & 4 & 23\\
      -5 & 1 & 26\\
      -7 & 0 & 27\\
     \hline
	\end{tabular}
	\caption[xxx]{\parbox[t]{0.8\textwidth}{
	Number of eigenvalues of $\tilde t^{(\frac{1}{2},s)}(u)$ 
		given by $\tilde \Lambda^{(\frac{1}{2},s) (\pm)}(u)$ for
	$N=3$, $s=1$. The total number of eigenvalues is $(2s+1)^{N}=3^{3}=27$.}
	}
       \label{table:n3s1}
     \end{table}
\begin{table}[htb] 
  \centering
  \begin{tabular}{|c|c|c|}\hline
      k & no. given by $\tilde \Lambda^{(\frac{1}{2},\frac{3}{2}) 
	(-)}(u)$ & no. given by 
	$\tilde \Lambda^{(\frac{1}{2},\frac{3}{2}) (+)}(u)$\\
    \hline
      7 & 16 & 0\\
      5 & 15 & 1 \\
      3 & 13 & 3\\
      1 & 10 & 6\\
      -1 & 6 & 10\\
      -3 & 3 & 13\\
      -5 & 1 & 15\\
      -7 & 0 & 16\\
     \hline
	\end{tabular}
	\caption[xxx]{\parbox[t]{0.8\textwidth}{
	Number of eigenvalues of $\tilde t^{(\frac{1}{2},s)}(u)$ 
		given by $\tilde \Lambda^{(\frac{1}{2},s) (\pm)}(u)$ for
	$N=2$, $s=3/2$. The total number of eigenvalues is $(2s+1)^{N}=4^{2}=16$.}
	}
       \label{table:n2s32}
     \end{table}
\begin{table}[htb] 
   \centering
   \begin{tabular}{|c|c|c|}\hline
       k & no. given by $\tilde \Lambda^{(\frac{1}{2},\frac{3}{2}) 
	 (-)}(u)$ & no. given by 
	 $\tilde \Lambda^{(\frac{1}{2},\frac{3}{2}) (+)}(u)$\\
	 \hline
	  10 & 64 & 0\\
	   8 & 63 & 1 \\
	   6 & 60 & 4\\
	   4 & 54 & 10\\
	   2 & 44 & 20\\
	   0 & 32 & 32\\
	   -2 & 20 & 44\\
	   -4 & 10 & 54\\
	   -6 & 4 & 60\\
	   -8 & 1 & 63\\
	  -10 & 0 & 64\\
	  \hline
	     \end{tabular}
	     \caption[xxx]{\parbox[t]{0.8\textwidth}{
	     Number of eigenvalues of $\tilde t^{(\frac{1}{2},s)}(u)$ 
		     given by $\tilde \Lambda^{(\frac{1}{2},s) (\pm)}(u)$ for
	     $N=3$, $s=3/2$. The total number of eigenvalues is $(2s+1)^{N}=4^{3}=64$.}
	     }
	    \label{table:n3s32}
\end{table}
\begin{table}[htb] 
	    \centering
	    \begin{tabular}{|c|c|}\hline
	     $E$ &  Bethe roots ${\tilde v}^{(-)}_{j}$\\
	      \hline
	      -8.78796 & 0.0781924 $\pm$ 0.150582 i \,, 0.573709 \\
	      -7.99601 & 0.377364+1.5708 i\,, 0.0718753 $\pm$ 0.150316 
	      i \\
	      -5.5443 & 0.191917 $\pm$ 0.145165 i \,, 0.529223 \\
	      -5.07143 & 0.375505+1.5708 i \,, 0.164075 $\pm$ 0.148506 
	      i\\
	      -4.31229 & 0.158193 $\pm$ 0.299905 i\,, 0.158448 \\
	      -3.36195 & 0.166008\,, 0.717455 $\pm$ 0.259354 i \\
	      -2.87198 & 0.358903+1.5708 i\,, 0.156172 \,, 0.784233 \\
	      -2.86245 & 0.33466 $\pm$ 0.286332 i \,, 0.337633 \\
	      -2.69332 & 0.371101+1.5708 i \,, 0.292713 $\pm$ 0.157296 
	      i \\
	      -2.31742 & 0.290731+1.5708 i\,, 0.617492+1.5708 i\,, 
	      0.146609\\
	      -1.52379 & 0.484424\,, 0.621449 $\pm$ 0.318594 i \\
	      -1.18428 & 0.356639+1.5708i\,, 0.464322\,, 0.659312 \\
	      -0.780678 & 0.288176+1.5708 i\,, 0.610874+1.5708 i\,, 
	      0.368261 \\
	      -0.379026 & 0.879352 $\pm$ 0.483137 i \,, 0.944398\\
	      -0.0249248 & 0.337585+1.5708 i\,, 0.934391 $\pm$ 0.266345 
	      i \\
	      0.389221 & 0.277491+1.5708i\,, 0.580415+1.5708 
	      i\,,0.97389\\
	      0.838091 & 0.245314+1.5708 i\,, 0.477481+1.5708 i\,, 
	      0.814847+1.5708 i \\
	       \hline
		  \end{tabular}
		  \caption[xxx]{\parbox[t]{0.8\textwidth}{
		  The 17 energies and corresponding Bethe roots 
		  given by $\tilde \Lambda^{(\frac{1}{2},1) (-)}(u)$ for  
		  $N=3\,, s=1\,, k=1\,, \eta=0.3i\,, 
		  \alpha_{-}=0.7i\,, \beta_{-}=0.2\,, 
		  \theta_{-}=0.5i\,, \alpha_{+}=1.2i\,, 
		  \beta_{+}=-0.2\,, \theta_{+}=-1.1i\,, 
		  \{ \epsilon_{i} \}=+1.$}
		  }
		 \label{table:energiesM}
\end{table}
\begin{table}[htb] 
	   \centering
	   \begin{tabular}{|c|c|}\hline
	    $E$ &  Bethe roots ${\tilde v}^{(+)}_{j}$\\
	     \hline
	     -9.55066 & 0.0900396 $\pm$ 0.151265 i \\
	     -5.71507 & 0.244797 $\pm$ 0.132886 i \\
	     -4.09573 & 1.50013 i\,, 0.182899 \\
	     -3.74447 & 0.786256 i\,, 0.174481 \\
	     -3.02558 & 0.514399 $\pm$ 0.220939 i\\
	     -2.07231 & 1.48515 i\,, 0.620007 \\
	     -1.79241 & 0.80571 i\,, 0.568604 \\
	     -1.1462 & 0.350408 +1.5708 i \,, 1.38463 i\\
	     -0.791024 & 0.093628+1.5708 i\,, 0.842728 i \\
	     -0.634944 & 0.979155 i\,, 0.863041 i\\
	      \hline
		 \end{tabular}
		 \caption[xxx]{\parbox[t]{0.8\textwidth}{
		 The 10 energies and corresponding Bethe roots 
		 given by $\tilde \Lambda^{(\frac{1}{2},1) (+)}(u)$ for  
		 $N=3\,, s=1\,, k=1\,, \eta=0.3i\,, 
		 \alpha_{-}=0.7i\,, \beta_{-}=0.2\,, 
		 \theta_{-}=0.5i\,, \alpha_{+}=1.2i\,, 
		 \beta_{+}=-0.2\,, \theta_{+}=-1.1i \,, 
		 \{ \epsilon_{i} \}=+1.$}
		 }
		\label{table:energiesP}
\end{table}

\end{document}